\documentclass[a4paper,11pt]{article}
\usepackage{pos}

\title{What is the correct definition of entropy for general relativistic field theory?}

\author*{Shuichi Yokoyama}

\affiliation{Department of Physical Sciences, College of Science and Engineering, \\
Ritsumeikan University,\\
Shiga, Japan}
\emailAdd{syr18046@fc.ritsumei.ac.jp}

\abstract{
Recently, the author and collaborators proposed a method to construct a new conserved charge different from the Noether one for general relativistic field theory on curved space-time with energy-momentum tensor covariantly conserved, and that the new conserved charge describes the entropy of the system. 
As concrete evidence of the proposals, it was shown that such a new conserved charge indeed exists for several classic gravitational systems, and that the proposed entropy density computed in them satisfies the local Euler's relation and the first law of thermodynamics concurrently and non-perturbatively with respect to the Newton constant. These developments are reviewed including brief discussions of physical implications of the local thermodynamic relations. 
}

\FullConference{The European Physical Society Conference on High Energy Physics (EPS-HEP2023)\\
 21-25 August 2023\\
Hamburg, Germany\\}

\begin{document}
\maketitle

\section{Introduction}
\label{Introduction}

The most innovative concept in thermodynamics is entropy.
This concept was created by Clausius, and its name was also by him combining the word 'energy' and a Greek word '$\rm{\tau\rho o\pi\acute\eta}$' to mean that it plays a role in describing energy or heat transport. (See \cite{cropper2004great}.) 
Indeed, entropy enables us to describe the laws of thermodynamics most concisely, which are basic tools to investigate a macroscopic system in thermodynamic equilibrium. 

Once a macroscopic system reaches thermodynamic equilibrium, thermodynamic observable quantities become stationary and uniform in the whole region of the system. 
This property for thermodynamic observables to be globally constant in space-time is in fact so strong that any macroscopic system does not go to thermodynamic equilibrium even after a long time passes. 
Even for such a system that does not reach thermodynamic equilibrium, thermodynamics is still useful if the system reaches {\it local} thermodynamic equilibrium, in which thermodynamic observables are constant in any subregion with the scale of the mean free path of constituent particles. 
Such local thermodynamic equilibrium systems can be seen in gravitational systems such as the homogeneous isotropic universe and a stable astronomical body realized on some curved space-time. Then it is natural to ask how the entropy is defined for such a local thermodynamic equilibrium system in curved space-time and how the laws of thermodynamics are modified therein. 

These questions are not easy to answer at all taking into account a long-standing problem on how to define the energy of a system in curved space-time. The problem is summarized as a trilemma that it is impossible to define a quantity in curved space-time respecting the conservation law, the gauge invariance of observables, and the consistent flat limit concurrently. 
The problem was already recognized by Einstein when he submitted a paper on general relativity \cite{https://doi.org/10.1002/andp.19163540702}. 
In order to maintain the conservation law of energy confirmed in the flat space-time, he introduced a quantity called pseudo-tensor to define energy in curved space-time. (See also \cite{Landau:1975pou}.)
The quantity was expected to describe the energy-momentum tensor of gravity. However, as suggested by the name, it is not a tensor. Therefore a quantity expected as energy defined by using a pseudo-tensor is not physical in gauge principle. In order to handle such a gauge-dependent property of the expected gravitational energy-momentum tensor, an idea is to evaluate it not in the bulk region but in the asymptotic boundary region. Such a quantity defined as a surface integral at constant time slice is called a quasi-local one. What is preferable in quasi-local energy is to achieve partial general coordinate invariance and to make it possible to evaluate masses of black holes without involving singularities, though it is accompanied by extra divergence even in the flat space-time, which needs to be regularized suitably by comparing a reference frame or adding quasi-local counter terms \cite{Arnowitt:1962hi,1962RSPSA.269...21B,Brown:1992br,Hawking:1995fd,Horowitz:1998ha,Balasubramanian:1999re,Ashtekar:1999jx}.
On the other hand, Komar pointed out that it is possible to construct a quantity combined with a vector field respecting both the conservation law and the gauge principle, which is called a Komar integral \cite{PhysRev.127.1411}. A Komar integral defined by using a suitably chosen time-like vector is called the Komar mass, which is expected as energy in general curved space-time. Although the Komar mass respects the conservation law and the gauge invariance, it does not reduce to the original definition of energy in the flat limit. 

As briefly overviewed above, how to define energy on curved space-time has been discussed for a long time and it has not reached one consensual definition yet. It goes without saying about the difficulty to give that of entropy on curved space-time.

\section{Proposals}

The author majored in theoretical particle physics in his Ph.D. course and has continued to investigate non-perturbative aspects of high energy physics and quantum field theory mainly to the end of resolving another long-standing problem of quantum gravity. 
During a project to realize a black hole including the effect of quantum gravity in a holographic approach \cite{Aoki:2020ztd}, the author recognized that quasi-local energy cannot precisely evaluate the energy of the gravitational system expected to realize a quantum black hole, and, trying to resolve this problem, he finally reached an expression of a volume integral for the purpose \cite{Aoki:2020prb}, which turned out to be essentially the same as written in an old textbook of general relativity by Fock \cite{Fock:1959} 
\begin{equation}
E= \int_{\Sigma_t} d^3x \sqrt{|g|} T^0\!_\mu n^\mu, 
\label{Energy}
\end{equation}
where $\Sigma_t$ is a hypersurface of constant time slice and $n^\mu$ is the time evolution vector field.  
This definition of energy achieves manifest general coordinate invariance and the consistent reduction in the flat limit, while it is not conserved unless the time evolution vector field is a Killing one. There are two important consequences of adopting the definition \eqref{Energy} \cite{Aoki:2020prb}. One is the agreement of the masses of well-known black holes computed by quasi-local energy. This was confirmed by carefully computing the contribution of the essential singularity expressed as a form of the delta function like the charge distribution of an electron in electromagnetism. The other is the violation of the equivalence principle between the inertial mass and the gravitational one for an extended object. Their deviation was explicitly computed for a spherically symmetric system in local equilibrium and described by gravitational self-interaction. The existence of such deviation is indeed consistent with that of tidal force as the non-removable one by changing an observer for any extended object. 

The definition of energy \eqref{Energy} can be extended to that of a general charge by replacing the time evolution vector field with a general one $v^\mu$ as 
$Q[v]= \int_{\Sigma_t} d^3x \sqrt{|g|} T^0\!_\mu v^\mu.$
Reaching this expression, one may ask oneself whether there is a case where this charge is conserved unless the vector field is a Killing one. 
The answer is yes if the energy-momentum tensor is covariantly conserved and there exists a vector field $\zeta^\mu$ to satisfy a differential equation such that
\begin{equation}
T^\mu\!_\nu\nabla_\mu \zeta^\nu=0.
\label{CTC}
\end{equation}
If the vector field $\zeta^\mu$ is not a Killing one, then the conserved charge $Q[\zeta]$ is not a Noether one but a new one that is not associated with any symmetry. One might wonder about its existence and its physical meaning. In fact, such a charge was found in a spherically symmetric gravitational system \cite{Kodama:1979vn}, and the vector field corresponding to $\zeta^\mu$ is referred to as the Kodama vector. The Kodama vector can be confirmed to satisfy the differential equation \eqref{CTC}.
It was shown in \cite{Aoki:2020nzm,Yokoyama:2023nld} that such a new conserved charge indeed exists explicitly in two gravitational systems different from the one in \cite{Kodama:1979vn}.
On the physical meaning, the author and collaborators proposed that the new conserved charge describes the entropy of the system \cite{Aoki:2020nzm}. 
An intuitive argument for this proposal is that entropy for a system governed by a fundamental theory such as a gravitational one must be conserved in general since dynamics in such a theory is reversible. Thus it is natural to expect that the new conserved charge plays the role.  
As concrete evidence, it was shown that both the local Euler's relation and the first law of thermodynamics hold non-perturbatively with respect to the Newton constant under the proposed interpretation in the two gravitational systems. These results are beriefly reviewed below. 

\section{Applications to local thermodynamic equilibrium systems}

\subsection{Homogeneous isotropic universe}

The first application is to the model of our universe \cite{Aoki:2020nzm}, which is approximately homogeneous, isotropic and in local thermodynamic equilibrium at each time slice on a very large scale beyond a cluster of galaxies. In this approximation, the metric is described by the Friedmann-Lema{\^i}tre-Robertson-Walker one, $g_{\mu\nu}dx^\mu dx^\nu = - dt^2 + a^2 ( \frac{dr^2}{1 - k r^2} + r^2 (d\theta^2 + \sin\theta^2 d\phi^2) )$, where $a$ is a function of time called the scale factor, $k=-1,0,1$ corresponding to open, flat, closed universe, respectively, and the energy-momentum tensor takes the form of a perfect fluid, $T^\mu\!_\nu=(p+\rho)u^\mu u_\nu+p\delta^\mu_\nu$, where $u^\mu$ is the fluid velocity and $\rho, p$ are the energy density, the pressure depending only on time, respectively. In the comoving frame, the Einstein equation reduces to   
$-8\pi G_N \rho = - 3(H^2 + \frac{k}{a^2}) +\Lambda, ~ 8\pi G_N p = ( 2\frac{dH}{dt} + 3H^2 + \frac{k}{a^2}  ) +\Lambda$,
where $G_N$ is the Newton constant, $H=\frac1a\frac{da}{dt}$ is the time-dependent Hubble parameter, and $\Lambda$ is the cosmological constant. 

Now solve the differential equation \eqref{CTC} to find a vector field $\zeta^\mu$ for a new conserved charge proposed as entropy. Although a general prescription thereof is not known yet, such a vector field is desired to satisfy the following conditions: one is to be generated from fluid velocity of the system, and the other is to respect the general covariance. These two conditions require the vector field to be proportional to the fluid velocity such that $\zeta^\mu=-\beta u^\mu$, where $\beta$ is an unknown scalar function of time. This ansatz reduces \eqref{CTC} to $\rho u^\mu\nabla_\mu \beta = pK \beta$, where $K=\nabla_\mu u^\mu$ is the so-called expansion. In the comoving frame, it can be solved as $\beta(t_1)=\beta_0 \exp(-\int_{t_0}^{t_1} dt pK/\rho)$, where $\beta_0=\beta(t_0)$ corresponds to an integration constant, and $K=3H$. 
Then the proposed entropy density $s$ is computed as $s=\sqrt{|g|} T^0\!_\mu \zeta^\mu=\sqrt{\tilde g}a^3 \rho \beta$, where $\tilde g=\frac{r^4}{1-kr^2}\sin^2\theta$. This can be rewritten as $s=\beta u$, where $v=\sqrt{\tilde g}a^3$ is the integrand of the volume at constant time slice, $u=\rho v$ is the internal energy density. This is regarded as the local Euler's relation $Ts=u$ by interpreting $\beta$ as the inverse local temperature, $\beta=1/T$. This interpretation is also consistent with the first law of thermodynamics, $T\frac{ds}{dt}= \frac{du}{dt}+p\frac{dv}{dt}$, which can be derived from a straightforward calculation. 
Employing the second Friedmann equation derived from the Einstein equation, $\frac{d\rho}{dt}+3H(\rho+p)=0$, one can show that the entropy density is conserved, as asserted.\footnote{The entropy density here is defined differently from the one in \cite{kolb1994early}. The latter is defined by the total entropy divided by the physical volume of the universe, and thus  is not conserved. The total entropy is conserved for both cases.} 

Therefore, it concludes that the energy is not conserved but the entropy is conserved as long as the universe is well approximated as homogeneous and isotropic.\footnote{Indeed, the total energy and the total entropy are computed as $U=\int d^3x u=V\rho$, $S=\int d^3x s=U/T$, where $V=\int d^3xv=\tilde V a^3$ is the volume of the space of the universe, or the physical volume, and $\tilde V=\int d^3x\sqrt{\tilde g}$ is the one labeled by the comoving coordinate system, or the comoving volume. For the closed universe of $k=1$, $\tilde V=2\pi^2$.}
This conclusion of the adiabatic evolution of the universe is not trivial at all if one takes into account the words of Clausius that 'The energy of the universe is constant, The entropy of the universe tends to a maximum,' \cite{cropper2004great} and the idea of Einstein to use pseudo-tensor to define energy for the purpose of respecting the conservation law \cite{https://doi.org/10.1002/andp.19163540702}. Another important consequence of local thermodynamics is the fate of the Big Bang of the universe regardless of any value of the Newton constant and any equation of state. This can be easily seen from the first law of thermodynamics, $s\frac{dT}{dt}=-p\frac{dv}{dt}$, which reads that if the volume decreases as time goes back, the temperature correspondingly increases since the entropy density and the pressure are both positive. 

\subsection{Spherically symmetric hydrostatic equilibrium}

The other application is to a hydrostatic equilibrium system with spherical symmetry \cite{Yokoyama:2023nld}. As previously, the energy-momentum tensor in such a system is described by a perfect fluid, but the energy density and the pressure are now functions only of the radial coordinate $r$. The metric in such a system is generally given by $g_{\mu\nu}\mathrm dx^\mu \mathrm dx^\nu= - f \mathrm dt^2 +h \mathrm dr^2 + r^2 (d\theta^2 + \sin\theta^2 d\phi^2) $, where $f,h$ are functions only of $r$. The balance between the attractive force of gravity and the repulsive one of matter is described by the Einstein equation, which is reduced in the comoving frame to $\frac{dp}{dr}= -\frac{(p+\rho)}{2}\frac{d\log f}{dr},$ $\frac{d\log h}{dr}=r h (8\pi G_N\rho +\Lambda) -\frac{\left(h-1\right)}{r},$ $\frac{d\log f}{dr}=r h (8\pi G_Np -\Lambda) +\frac{\left(h-1\right)}{r}$. Substituting the 3rd equation into the 1st one, one obtains the Tolman-Oppenheimer-Volkov equation \cite{PhysRev.55.374}. 

Now solve the differential equation \eqref{CTC} and find a vector field $\zeta^\mu$ for a conserved entropy. To the end, as previously, the vector field is set to be proportional to the fluid velocity $\zeta^\mu=-\zeta u^\mu$, where $\zeta$ is an unknown scalar function of $r$. This ansatz reduces \eqref{CTC} to $\frac{d\zeta}{dr} = -\frac{p}{\rho }\frac{ \zeta }{\rho + p}\frac{d\rho}{dr}$ in the comoving frame, which can be solved as $\zeta =\beta_0 u^t f (1 + \frac p\rho)$ with $\beta_0$ an integration constant. 
Substituting this solution into the proposed definition of entropy density, one obtains $s=\beta_0 f^{1/2} (u + pv) $, where $v =r^{2} \sqrt{h}\sin\theta$ is the integrand of the volume at constant time slice, $u=\rho v$ is the internal energy density. Therefore, if $\beta_0 f^{1/2} =:1/T$ is interpreted as the inverse local temperature, then the local Euler's relation is obtained as $Ts=u+pv$. Importantly, this interpretation of the local temperature exactly matches the one derived by Tolman in a different way \cite{PhysRev.35.904}. (See also \cite{alma991013077249704706,Misner:1974qy}.) This local temperature leads to a differential equation $\frac{dT}{dr}= \frac{T}{p+\rho}\frac{dp}{dr}$ and also to the first law of thermodynamics, $T\frac{ds}{dr}= \frac{du}{dr}+p\frac{dv}{dr}$. Note that the associated entropy current is $s^\mu=\beta_0 u^tf^{3/2}(u+pv)u^\mu$ and its conservation $\partial_\mu s^\mu=0$ can be shown in a radially moving frame.

It concludes that a spherically symmetric hydrostatic equilibrium system such as a stable spherical star obeys the local thermodynamic relations non-perturbatively in the Newton constant. 

\section*{Acknowledgement}
The author would like to thank S.~Aoki and T.~Onogi for collaboration. He is also grateful to colleagues and peer reviewers for giving constructive comments. The author is supported by the Grant-in-Aid of the Japanese Ministry of Education, Sciences and Technology, Sports and Culture (MEXT) for Scientific Research (No.~JP22K03596). 

\bibliographystyle{JHEP}
\bibliography{stellar.bib}

\end{document}